\documentclass[conference]{IEEEtran}
\usepackage{cite}

\usepackage{bm}

\ifCLASSINFOpdf
   \usepackage[pdftex]{graphicx}
   \graphicspath{{figures/}}
\else
\fi

\usepackage{amsmath}
\usepackage{fixltx2e}

\usepackage{subfig}

\usepackage{balance}

\usepackage{stfloats}

\usepackage{amsmath}

\usepackage{graphicx}

\usepackage{amssymb}

\usepackage{makecell}

\usepackage{enumitem}
\setlist[itemize]{noitemsep, topsep=0pt}

\usepackage{mathtools}

\usepackage{algorithm,algorithmic}

\begin{document}

\title{Realtime Scheduling and Power Allocation Using Deep Neural Networks}
\author{
\IEEEauthorblockN{Shenghe Xu, Pei Liu, Ran Wang and Shivendra S. Panwar}
\IEEEauthorblockA{Department of Electrical and Computer Engineering, NYU Tandon School of Engineering, Brooklyn, NY, USA
\\\{shenghexu, peiliu, rw1691, panwar\}@nyu.edu}

}

\maketitle



\begin{abstract}
With the increasing number of base stations (BSs) and network densification in 5G, interference management using link scheduling and power control are vital for better utilization of radio resources. However, the complexity of solving link scheduling and the power control problem grows exponentially with the number of BS. Due to high computation time, previous methods are useful for research purposes but impractical for real time usage.  In this paper we propose to use deep neural networks (DNNs) to approximate optimal link scheduling and power control for the case with multiple small cells. A deep Q-network (DQN) estimates a suitable schedule, then a DNN allocates power for the corresponding schedule. Simulation results show that the proposed method achieves over five orders of magnitude speed-up with less than nine percent performance loss, making real time usage practical.
\end{abstract}

\section{Introduction}
Network densification has been proposed as one of the main schemes to meet the increasing traffic demand in 5G \cite{bhushan2014network}. With higher density of BS and limited sub-6Ghz spectrum resources, nearby BSs may need to utilize the same frequency band to serve users. However, utilization of the same frequency band could cause higher interference and bring considerable degradation of network performance. Interference management becomes important for better utilization of spectrum resources. But for BSs in a certain area, the number of possible schedules of the system grows exponentially with the number of BSs. In addition, suitable power allocations have to be found for each schedule to provide the optimal network performance. Though exhaustive search combined with Geometrical Programming (GP) could provide near optimal solutions \cite{chiang2007power}, high computation time makes this scheme impractical for real time usage. 

In recent years, deep neural networks (DNNs)  \cite{lecun2015deep}  have shown promising results for various tasks including image recognition \cite{hinton2006reducing}, speech recognition \cite{hinton2012deep} and function approximation \cite{liang2016deep}. Several papers have also applied neural networks to the research of wireless communications \cite{mennes2018neural, o2017deep,karanov2018end, zhang2018deep}. Though previous papers \cite{lee2018deep, liang2018towards} have considered using DNNs for power allocation, no one has considered using DNNs for both link scheduling and power allocation with multiple interfering small cells. 

On the other hand, reinforcement learning \cite{sutton1998reinforcement},\cite{riedmiller2005neural}, especially deep reinforcement learning has achieved state-of-the-art results on various applications such as playing Atari games \cite{mnih2015human}, indoor navigation \cite{zhu2017target}, and various other tasks \cite{duan2016benchmarking}. Several papers have also proposed to use deep reinforcement learning to solve problems in wireless communications \cite{he2017optimization, yu2018deep}. The problem of scheduling in wireless networks also require the agent to pick the best policy based on the environment. It shows great resemblance to the problems solved by deep reinforcement learning. 

In this paper we propose to use a DQN and a DNN for link scheduling and power allocation in a multi-cell scenario, respectively. We consider multiple small cells using the same channel resources, where each cell could schedule an uplink or a downlink transmission for one of the users at a time slot. 

The main contributions of this paper are:
\begin{itemize}
   \item An efficient power allocation method using a DNN is proposed; the method can dynamically allocate power for BSs and UEs to maximize the weighted sum rate of the whole system. 
   \item We propose a link scheduling scheme using a DQN. Trained with data generated from the power allocation neural network, the link scheduling network could provide suitable schedules combined with either GP based power control or neural network based power control. 
   \item Simulation results show that compared with exhaustive search and GP, the performance loss of weighted sum-rate (WSR) achieved by joint scheduling and power allocation using DQN and DNN is 8.66\%, with over five orders of magnitude times speed-up.  Exhaustive search combined with power allocation neural network could provide over four orders of magnitude times speedup with a 5.71\% performance loss. The scheduling DQN combined with GP could also provide four orders of magnitude times speed-up with 6.12\% performance loss. 
\end{itemize}

The rest of this paper is organized as follows. Section \ref{Sec: system} provides the system description and problem formulation. In Section \ref{Sec: scheduler} the scheduling and power allocation algorithm is presented. The simulation setup is included in Section \ref{Sec: simulation}. In Section \ref{Sec: evaluation} we evaluate the performance of our method. Conclusions are drawn in Section \ref{Sec: conclusion}.

\section{System Model and Problem Formulation}\label{Sec: system}
In this paper we consider $N$ small cell BSs using the same channel resources to serve several UEs. Each BS is associated with $M$ UEs. The BSs are backhauled with fiber or orthogonal wireless links. The downlink transmission for UE $i$ is associated with weight $w_{2i}$, while its uplink transmission is associated with weight $w_{2i+1}$. The weights may be assigned dynamically at every slot according to the back-pressure method \cite{tassiulas1992stability} or any other scheduling scheme.  An example of the system is shown in Fig. 1. 
\begin{figure}[!t]
	\centering
	\includegraphics[width=.4\textwidth]{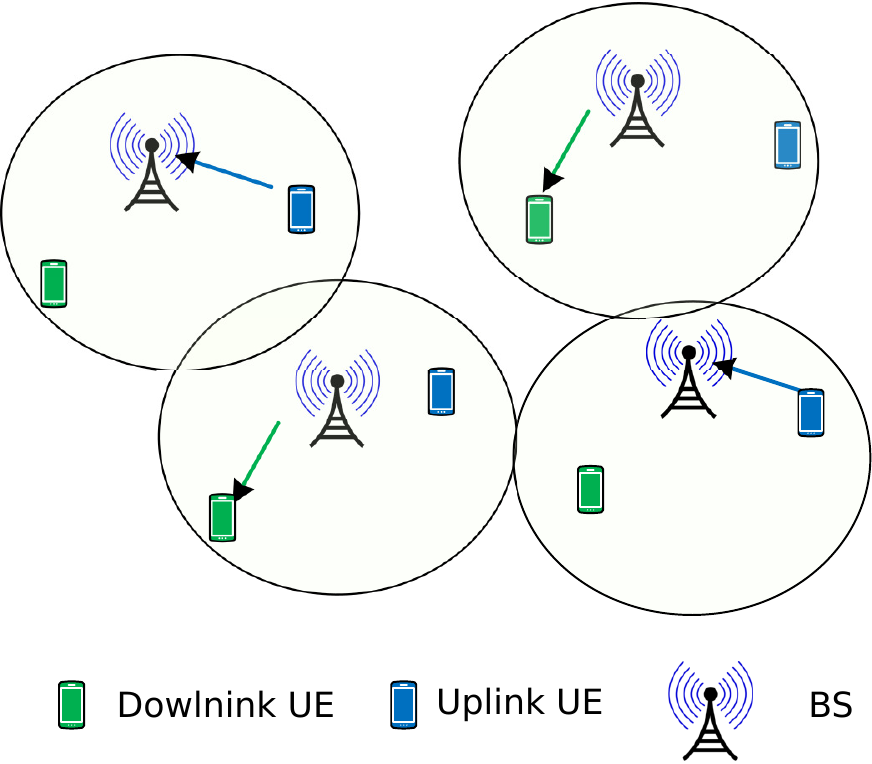}
\caption{System Example}
\label{Fig: Topology}
\end{figure}
 
The objective of the system is to maximize the WSR of all the small cells. The weighted sum-rate maximization (WSRMax) problem has been studied in several previous papers\cite{weeraddana2012weighted, zhang2011weighted, gjendemsjo2008binary} . The general form of the WSR maximization problem is NP-hard\cite{luo2008dynamic}.  In this paper we focus on the scenario where multiple small cell BSs are communicating with several associated users. Each user is associated with only one BS. Each BS could choose to schedule either a downlink or uplink transmission for one of the UEs. For link $l_i$, the signal-to-interference-plus-noise ratio (SINR) can be written as 
\begin{equation}
  SINR_i = \frac {  | h_{i,i} |^2 p_i }{ \sum_{j \neq i} | h_{i,j} |^2 p_j  + \mathcal N_i} ),
\end{equation}
where $p_i$  is the transmit power for link $i$, $h_{i,j}$ is the complex channel response between the transmitter of link $l_j$ and receiver of link $l_i$,.

Then for all the links $l_i$ and weights $w_i$, $i = 1,2, ... , 2MN$, the problem can be formulated as 

\begin{equation}
\max_{p_1,p_2,..., p_N }  \sum_{i=1}^{N} w_i W \log (1 + \frac {  | h_{i,i} |^2 p_i }{ \sum_{j \neq i} | h_{i,j} |^2 p_j  +  \mathcal N_i} ), 
\end{equation}
\begin{equation}
\nonumber \text{s.t. }  0 \leq p_i \leq P_i^{max},   \forall i = 1,2, ... , 2MN.
\end{equation}
where W is the channel bandwidth, $\mathcal N_i$ is the background noise at the receiver of link $l_i$, $P_i^{max}$ is the maximum transmission power for link $l_i$. Since for each small cell, only one downlink or uplink could be scheduled, there are $2^NM^N$ possible schedules, where each schedule corresponds to a power allocation problem with $N$ links. 

The optimization problem in (2) is a nonlinear nonconvex problem. But GP could be used to get a near-optimal solution of this problem. The problem can be first written as a signomial programming (SP) problem. Then according to \cite{chiang2007power}, an iterative procedure could be used to solve this problem by constructing a series of GPs, each of which could be easily solved. This procedure is provably convergent and the optimal power allocation could almost always be obtained using this method. For more details regarding this procedure, we refer to \cite{chiang2007power}.

\section{Link Scheduling and Power Control Using Deep Neural Networks}\label{Sec: scheduler}
\subsection{The Power Allocation Deep Neural Network}
Since the possible schedules for WSRMax problem grows exponentially with the number of small cells, it is impossible to exhaustively search for the best schedule for real time usage. We propose to use neural networks for power allocation and link scheduling. In this paper we use a fully connected neural network to solve the power allocation part of the WSRMax problem. The network architecture is shown in Fig. 2 (a). The matrix $\bm{G}$ containing all the channel gains $| h_{i,j} |^2$ related the corresponding schedule is first processed by a layer of size 64. The vector $\bm{w}$ containing all the weights is processed by another layer with 32 neurons. The vector $\bm{u}$ indicating uplink as 0 and downlink as 1 is handled by a layer of size 32. Then the three layers are concatenated together and processed by three fully connected hidden layers each with size 256, 128 and 64. The final layer of size 4 outputs the vector $\bm{p}$ containing all the power allocations.

\begin{figure}[!ht]
		\centering
		\subfloat[Power Allocation Network\label{subfig:power}]{
			\includegraphics[width=.23\textwidth]{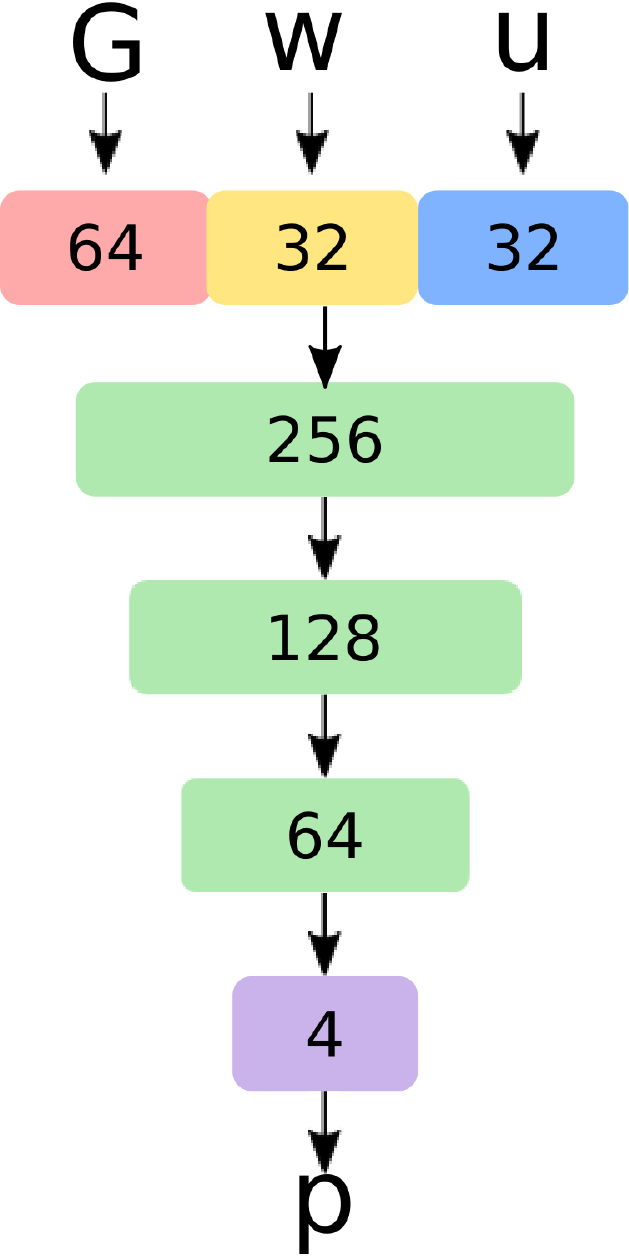}}
		\subfloat[Scheduling Network\label{subfig:schedule}]{
			\includegraphics[width=.1925\textwidth]{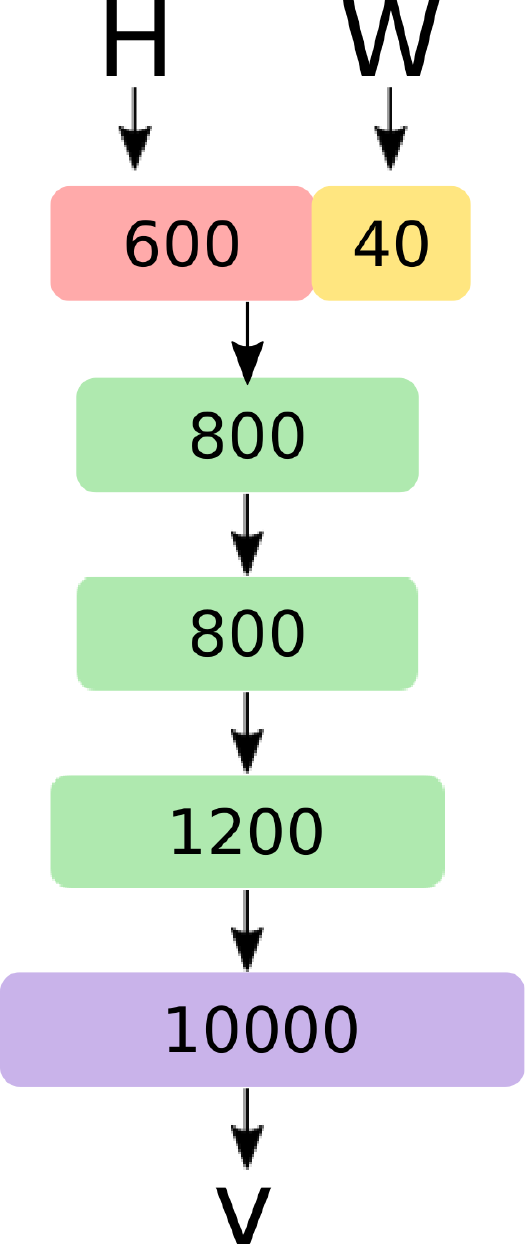}}
	
	\caption{Network Structures}	
\label{Fig: Network}
\end{figure}

The network is trained with power allocation results from GP. The powers $p_i$ are divided by $P_i^{max}$ so that all output should be in the range from 0 to 1. To approximate the solution from GP, MSE is used as the loss function to train the neural network. 

Since neural networks could process data in batches, $\bm{G}$, $\bm{w}$ and $\bm{u}$ from all possible schedules could be combined into one or a few batches and processed together. All the corresponding power allocations could be obtained at the same time. Then the weighted sum rate could be computed for each of the schedules. A very simple scheduling scheme is to pick the schedule with the highest weighted sum rate. 

\subsection{The Scheduling Deep Neural Network}
Though simulation results show that exhaustive search over all the WSR provided by the power allocation network could provide 24967 times speed-up with less than 6\% performance loss, the number of schedules grows exponentially with the number of BS. Computation time would increase greatly with the number of users and BSs. We propose to use a DQN to provide an estimation of the best user selection and uplink/downlink scheduling. Instead of learning the Q function from the environment, the DQN learns from the WSRs achieved by the power allocation DNN. The network structure of the DQN is shown in Fig. 2 (b). The channel matrix $\bm{H}$ containing $| h_{i,j} |^2$ between all pairs of transmitter and receivers and the weight matrix $\bm{W}$ are each fed into input layers with 600 and 40 neurons. In addition, the network has three hidden layers with size 800, 800, 1200 and an output layer with size 10000. Each of the outputs correspond to the value function, or in this case, the WSR for a schedule. 

Since the DQN should pick the best possible schedule combined with the power control network, it is trained with WSR provided by the power control DNN instead of GP. MSE is used as loss function to train the DQN to give an estimation for WSR of each schedule. However, simulation results show that the DQN could also provide promising performance combined with power control with GP. Though the WSRs provided by the DNN are only approximations of the WSRs provided by GP, the DQN is still able to estimate the performance of GP by learning from the DNN.

\section{Simulation setup}\label{Sec: simulation}
To evaluate the performance of the proposed methods, we consider four small cells randomly distributed in a 120 meter by 120 meter square area. The minimum distance between each pair of the BSs is 40 meters. UEs are randomly distributed in a disc area around its associated BS with a minimum distance of 10 meters and a maximum distance of 40 meters. Though the coverage area of two BSs may overlap, users are associated with the BS with the highest channel gain. Each BS serves five users.

We adopt the channel model from 3GPP\cite{3gpp.36.828}.  The probability of LOS for the channels can be found in \cite{3gpp.36.828}. Other simulation details are shown in Table 1. The spectral efficiency is capped at 7 bits/sec/Hz to match the peak spectrum efficiency of a system with practical modulation and coding. 
\begin{table}
\renewcommand{\arraystretch}{1.3}
\caption{Simulation Parameters. SSD: Shadowing standard deviation. R is in kilometers. }
\label{tab:table1}
\begin{center}
	\begin{tabular}{| c | c |}
	 \hline
	 Parameter & Value \\
	 \hline 
	 System bandwidth & 10 MHz \\
	 \hline
	 SSD between BS and UE & LOS: 3 dB, NLOS: 4 dB \\
	 \hline
	 BS to UE path loss &  \makecell{ LOS: $PL(R) = 103.8 + 20.9 \log_{10}(R)$\\NLOS: $PL(R) = 145.4 + 37.5 \log_{10} (R)$}\\
	 \hline
	 UE to UE path loss & \makecell{ If $R \leq 50m, PL(R) = 98.45$ \\ $ + 20\log_{10}(R)$. \\Else, $PL(R) = 55.78 + 40\log_{10}(R)$} \\
	 \hline
	 Maximum Power & BS: 24dBm, UE: 23dBm \\
	 \hline
	 Noise Figure & BS: 12dB, UE: 9dB \\
	
	 \hline

	 \end{tabular}

\end{center}
\end{table}
The location of both BSs and UEs are randomly generated each time. Weights are also randomly generated from the uniform random distribution between 0 and 1 for each topology. 

For the power allocation network, 900 sample topologies are used for training, 100 are used for validation, and each sample corresponds to 10000 schedules. So the total number of training data is 9,000,000. The plot of training loss, validation loss and the validation mean WSR over the training time of the DNN is shown in Fig. 3. As the training and validation MSE drops with time, the DNN learns from the power allocation obtained from GP, so the validation mean WSR increases gradually.
\begin{figure}[!t]
	\centering
	\includegraphics[width=.47\textwidth]{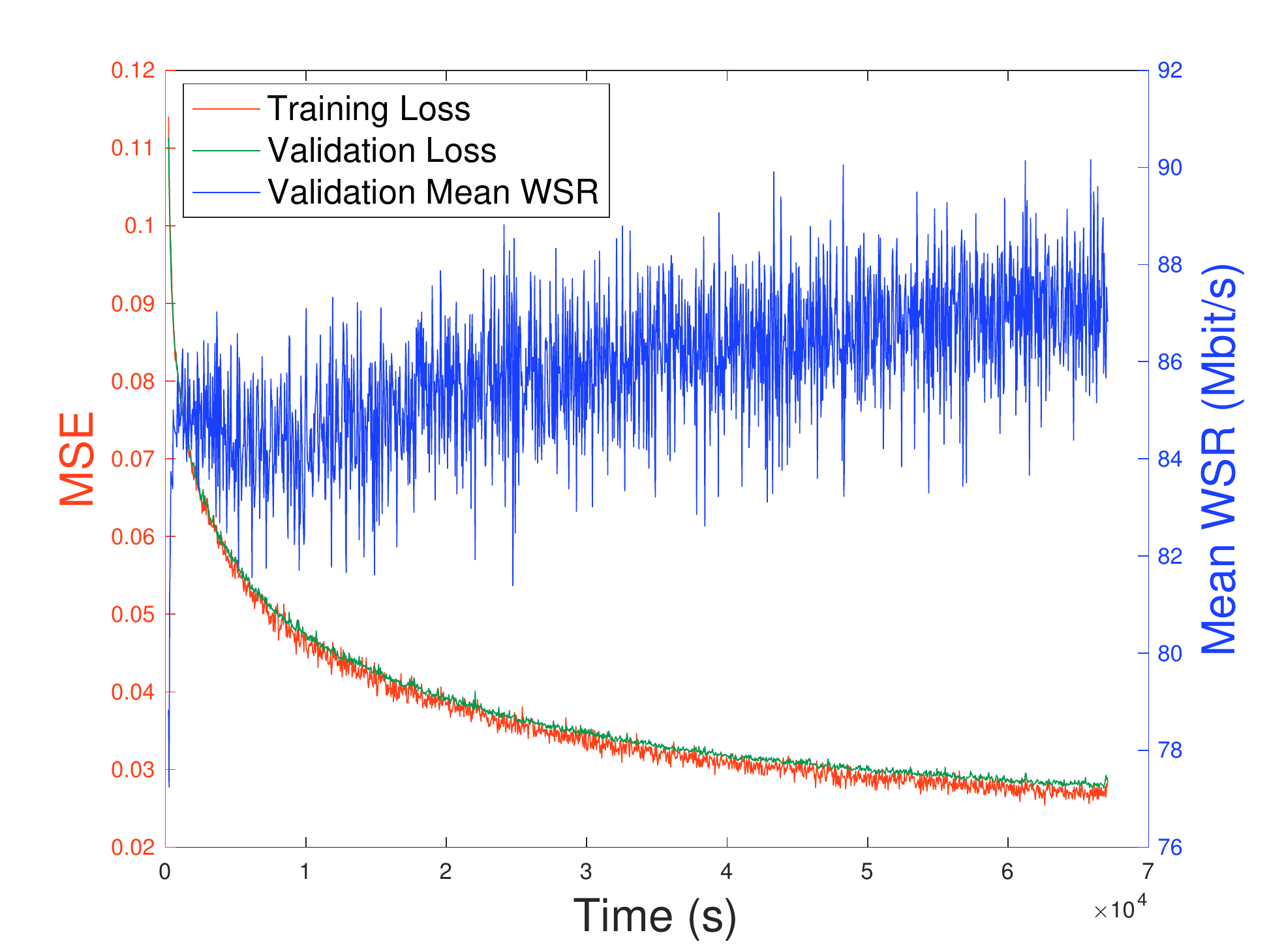}
\caption{MSE and Mean WSR over DNN Training Time}
\label{Fig: DNNTrainingVsTime}
\end{figure}

After the training of the power allocation network completes, the scheduling DQN is trained using the WSRs obtained by the power allocation DNN. 870,000 random samples with different topologies are used for training of the DQN. 30,000 samples are used for validation. The plot of training loss, validation loss and the validation mean WSR over training time of the DQN is shown in Fig. 4. As the MSE drops, the DQN learns to estimate the WSR for each schedule, and the mean of maximum WSR for each validation topology increases. 
\begin{figure}[!t]
	\centering
	\includegraphics[width=.47\textwidth]{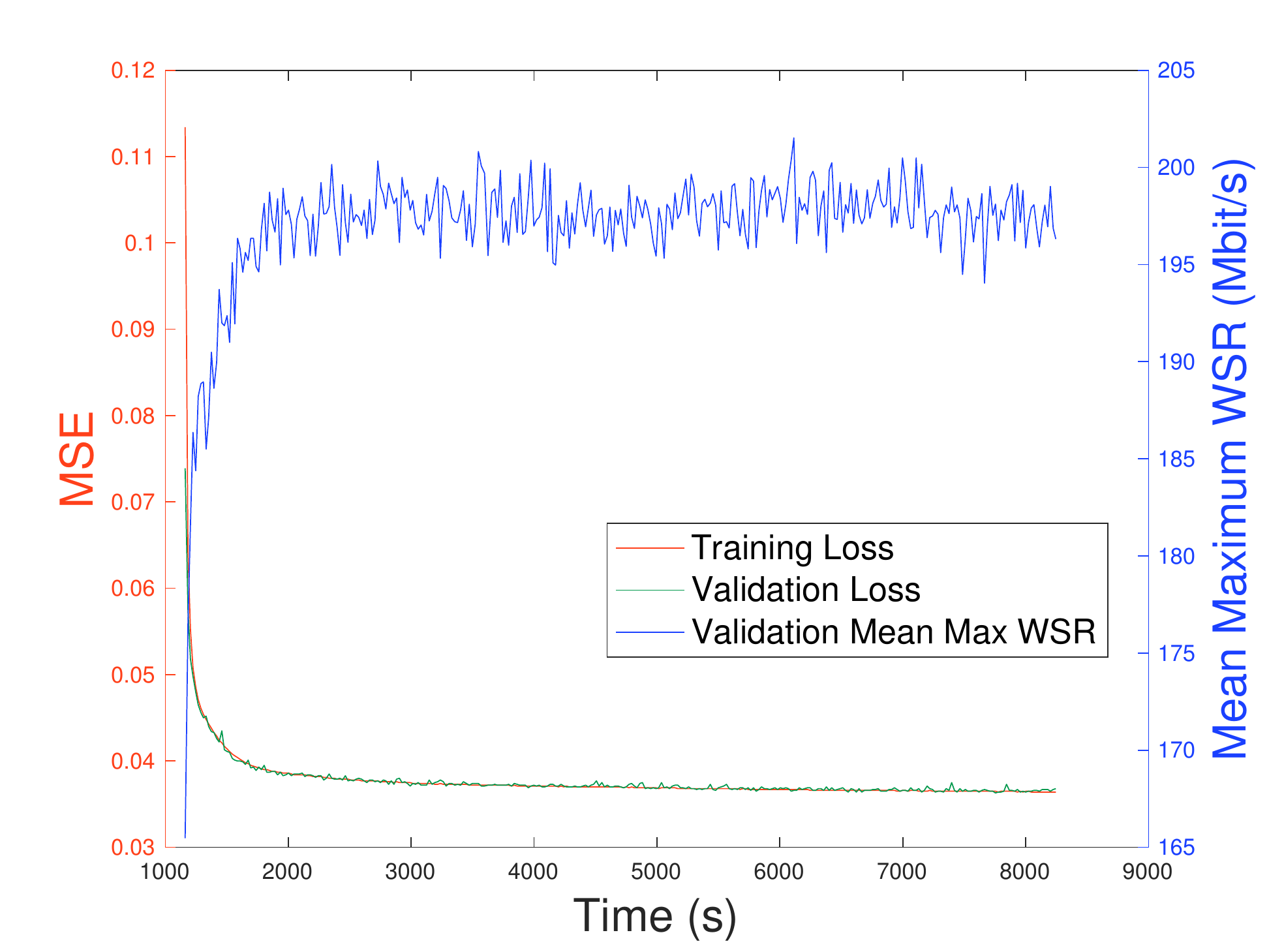}
\caption{MSE and Mean Maximum WSR over DQN Training Time}
\label{Fig: DQNTrainingVsTime}
\end{figure}

\section{Performance Evaluation}\label{Sec: evaluation} 
We compare our proposed scheduling scheme with three other methods. For the baseline method, links are randomly selected for each cell, then powers are allocated using GP; this method is denoted as Random-GP. For the second baseline method, in each of the small cell, the link with the highest weight is selected, power allocations are obtained from GP, we refer to this method as Greedy-GP. The method that exhaustively searches for the best schedule using GP is referred as Exhaustive-GP. For the method that computes power allocation for all possible schedules with DNN and picks the best, we denote it as Max-DNN. The scheme that uses schedules provided by the DQN and power allocation obtained from GP is DQN-GP. The method that uses both DQN and DNN is referred as DQN-DNN. To further improve the estimation accuracy of DQN, we propose to pick the top five schedules obtained from DQN and run DNN for each of the schedules. The schedule with the highest WSR is picked, we denote this method as DQN-DNN-5. We also add a baseline method where each BS picks the link with the highest weight and transmit with its maximum power, this method is denoted as Greedy-MP. 

All the methods are evaluated for 10000 samples, each with different BS and UE locations. The cumulative distribution function of the WSRs achieved by each of the methods is shown in Fig. 5. 
\begin{figure}[!t]
	\centering
	\includegraphics[width=.47\textwidth]{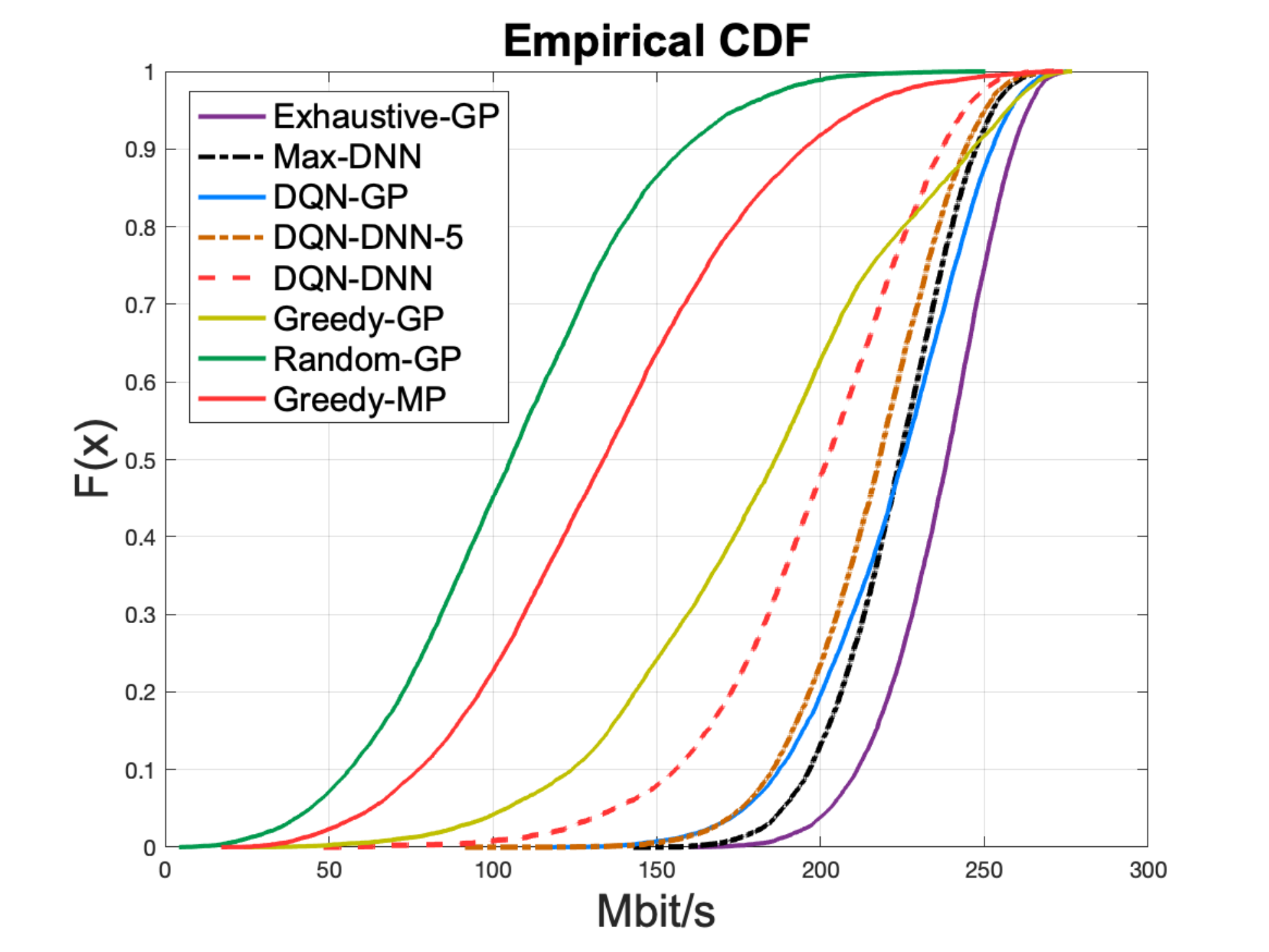}
\caption{Empirical CDF Comparison}
\label{Fig: Empirical CDF Comparison}
\end{figure}
The mean WSR for each method is shown in Table 2. The last column shows the performance loss compared with Exhaustive-GP.

For running time comparisons, all the experiments were run on servers allocated with one core of an Intel Xeon E5-2690 v4 CPU and 4GB memory. Though GPUs are used for training of the DNN and DQN, only a CPU is used for running time comparisons. For Exhaustive-GP, Random-GP and Greedy-GP, the running time was averaged over 1000 samples. The code for GP was written in Matlab, run with Matlab 2017a. GGPLAB was used to solve GP \cite{mutapcic2006ggplab}. Only the time spent to solve GP was included for CPU time evaluation.  According to the Matlab runtime profiler, over 90 percent of the time was spent on executing functions of GGPLAB. For DNN and DQN, the program was written in Python 2.7 with Tensorflow 1.5.0. The power scheduling DNN running time was averaged over 10000 samples; the average inference time for one sample is 0.38ms. For Max-DNN, power allocation is obtained for all the 10000 schedules of one topology in one batch. Since computation for each set of power allocations within a batch is parallel, using 2 cores of the CPU could roughly bring two times speed-up for Max-DNN. The mean WSR (MWSR) and running time are shown in Table 2. 

\begin{table}
\renewcommand{\arraystretch}{1.0}
\caption{Performance Comparison}
\label{tab:table2}
\begin{center}
	\begin{tabular}{ c || c  c  c  c  c|}
		\hline 
		Method             & MWSR 			& CPU time 		& MWSR Loss \\
		\hline 
		Exhaustive-GP 	& 236.45 Mbit/s		& 1615.34s 		& 0\% 			\\
		\hline
		Max-DNN		& 222.95 Mbit/s 	& 0.0647s			& 5.71\%			\\
		\hline
		DQN-GP 		& 221.99 Mbit/s		& 0.1350s			& 6.11\%			\\
		\hline
		DQN-DNN-5 	& 215.96 Mbit/s		& 0.0090s			& 8.66\%			\\ 
		\hline
		DQN-DNN 	& 198.24 Mbit/s		& 0.0074s			& 16.16\%			\\ 
		\hline
		Greedy-GP	& 183.65 Mbit/s		& 0.1827s			& 22.33\%			\\	
		\hline
		Random-GP 	& 107.28 Mbit/s		& 0.2132s 	 	& 54.63\% 		\\
		
		\hline
		
		\end{tabular}
\end{center}
\end{table}

It can be seen from table 2 that with DQN-DNN-5, with the settings in our experiment, we can achieve over 170000 times speedup with less than nine percent performance loss. The DQN trained from the rates achieved by power allocation DNN could also provide suitable schedules for GP. Without power allocation DNN, training the DQN with GP provided rates could be too time consuming to be implemented. It is interesting to see that though Random-GP, Greedy-GP and  DQN-GP all involves power control with GP, there is considerable difference between running time of these three methods. In fact, on average the power allocation problem for schedules in Random-GP takes 16.77 iterations to solve, power allocation for schedules selected by Greedy-GP requires an average of 14.41 iterations to solve, while on average the schedules selected by the DQN only require 8.77 iterations for GP to provide the power allocation results. The schedules selected by the DQN requires less computation time for power allocation, but can still achieve similar performance as exhaustive search.

\section{Conclusion and Future work}\label{Sec: conclusion}
In this paper, to solve the problem of WSRMax, we proposed to use a DQN to replace exhaustive search for scheduling and a DNN for power allocation. Simulation results show that using the DQN and DNN could bring five orders of magnitude of computation time reduction with less than nine percent performance loss. Future work includes finding suitable methods to train the DQN and DNN together. In addition, a scheme that could directly use neural networks to optimize WSR is yet to be devised. 

\section{Acknowledgements}
This work was supported by the National Science Foundation under Grant No. 1527750, \balance as well as the NYSTAR Center for Advanced Technology in Telecommunications (CATT), and NYU WIRELESS.



\bibliographystyle{IEEEtran}
\bibliography{IEEEabrv,DL_scheduling}
\end{document}